\documentclass[3p,twocolumn,longtitle,reversenotenum,sort&compress,lefttitle]{elsarticle}

\pdfoutput=1

\usepackage[utf8]{inputenc}
\usepackage[T1]{fontenc}

\DeclareUnicodeCharacter{03B2}{\ensuremath{\beta}}
\DeclareUnicodeCharacter{03C1}{\ensuremath{\rho}}
\DeclareUnicodeCharacter{03C3}{\ensuremath{\sigma}}
\DeclareUnicodeCharacter{2264}{\ensuremath{\leq}}
\DeclareUnicodeCharacter{2265}{\ensuremath{\geq}}

\PassOptionsToPackage{hyphens}{url}\usepackage[hidelinks]{hyperref}

\usepackage{libertine}
\usepackage{microtype}

\usepackage[version=4]{mhchem}

\usepackage{enumitem,amssymb}
\usepackage{tabularx,booktabs}

\makeatletter
\let\@afterindenttrue\@afterindentfalse
\makeatother

\nopreprintlinetrue

\begin{document}
\begin{frontmatter}

\title{Untargeted analysis of volatile markers of post-exercise fat oxidation\\in exhaled breath}

\author[aff1]{André~Homeyer}
\author[aff2]{Júlia~Blanka~Sziládi} 
\author[aff1]{Jan-Philipp~Redlich} 
\author[aff2]{Jonathan~Beauchamp}
\author[aff2]{Y~Lan~Pham\corref{corresp}}
\ead{y.lan.pham@ivv.fraunhofer.de}

\address[aff1]{Fraunhofer Institute for Digital Medicine MEVIS, Bremen, Germany}
\address[aff2]{Department of Sensory Analytics and Technologies, Fraunhofer Institute for Process Engineering and Packaging IVV, Freising, Germany}

\cortext[corresp]{Corresponding author}

\begin{abstract}
Breath acetone represents a promising non-invasive biomarker for monitoring fat oxidation during exercise. However, its utility is limited by confounding factors, as well as by the fact that significant changes in concentration occur only hours post-exercise, which makes real-time assessment difficult. We performed an untargeted screening for volatile organic compounds (VOCs) that could serve as markers of fat oxidation beyond acetone, and investigated whether breath measurements taken during exercise could predict post-exercise changes in fat oxidation. Nineteen participants completed two 25-min cycling sessions separated by a brief 5-min rest period. VOC emissions were analysed using proton-transfer-reaction time-of-flight mass spectrometry (PTR-TOF-MS) during exercise and after a 90-min recovery period. Blood β-hydroxybutyrate (BOHB) concentrations served as the reference marker for fat oxidation. Among 773 relevant analytical features detected in the PTR-TOF-MS measurements, only four signals exhibited strong correlations with BOHB (ρ~≥~0.82, p~=~0.0002)---all attributable to acetone or its isotopologues or fragments. End-of-exercise measurements of these signals enabled accurate prediction of participants with substantial post-exercise BOHB changes (F1~score~≥~0.83, accuracy~=~0.89). Our study did not reveal any novel breath-based biomarkers of fat oxidation, but it confirmed acetone as the key marker. Moreover, our findings suggest that breath acetone measurements during exercise may already enable basic predictions of post-exercise fat oxidation.


\end{abstract}

\end{frontmatter}

\section{Introduction}

Overweight and obesity are associated with more than 1.2 million deaths annually in Europe and surrounding regions; on average, 63\% of men and 54\% of women in the EU are overweight, and the trend is rising~\citep{who2022obesity}. Physical activity is one of the cornerstones in reducing overweight and obesity~\citep{reiner2013}. One popular method is training on an ergometer, both in the gym and at home. A variety of factors influence fat burning, including training intensity and duration, diet, fitness level and time of day~\citep{purdom2018, amarogahete2019, achten2004}. Therefore, it would be desirable to receive feedback on fat burning status during training in order to monitor exercise efficacy.

To date, there is no convenient method of monitoring the state of fat burning during exercise. Most ergometers are equipped with a calorie counter that roughly estimates the current energy consumption based on power measurements and other data~\citep{pageglave2018}. Modern fitness wristbands estimate energy consumption based on heart rate~\citep{shcherbina2017}, but this does not directly translate to fat burning, as the body first uses up glycogen reserves before it burns fat.

Fat burning, technically referred to as "fat oxidation", is commonly quantified based on three ketone bodies: acetoacetate, β-hydroxybutyrate (BOHB) and acetone~\citep{anderson2015, laffel1999}. These molecules are produced in the liver as metabolites of free fatty acids. When carbohydrate availability is reduced, for instance during fasting or prolonged exercise, systemic ketone body concentrations increase~\citep{laffel1999} and are therefore exploitable as biomarkers of fat oxidation~\citep{delorbe2022}. Ketosis represents a state of elevated ketone levels~\citep{laffel1999}.

These ketone bodies can be quantified using different methods. Test strips for detecting elevated acetoacetate levels in excreted urine have existed for many years~\citep{suntrup2020}. Point-of-care devices are available for quantifying BOHB levels in blood via capillary blood sampling from the fingertip or via subcutaneously implanted glucose sensors adapted for ketone monitoring~\citep{abegg2020}. These tools are routinely employed to assess diet‑induced fat oxidation and nutritional ketosis~\citep{laffel1999}. Because they require repeated finger‑pricks or continuous sensor wear, however, they are poorly suited to assessing exercise‑induced fat oxidation~\citep{delorbe2022}.

Acetone is the only ketone body that is volatile and can diffuse freely from the blood into the airways and thereby manifest in exhaled breath~\citep{anderson2015}. This makes acetone an attractive target for non‑invasive monitoring of fat oxidation via breath. Multiple studies have demonstrated strong, non-linear correlations between breath acetone (BrAce) concentrations and BOHB levels~\citep{anderson2015, suntrup2020, guentner2017}, prompting significant research interest in developing BrAce sensors~\citep{lee2024, delorbe2022, weber2021, kim2020, guentner2017}.

One challenge that could limit the usefulness of BrAce as a biomarker for fat oxidation during exercise is its variability~\citep{drabinska2021, ruzsanyi2017}. Baseline concentrations can vary widely among individuals, ranging from 100 to 2500 parts per billion, by volume, and levels can fluctuate within individuals over hours or days~\citep{spanel2020, anderson2015}. This variability arises because acetone is produced and metabolized through several pathways influenced by numerous factors~\citep{kalapos2003, ruzsanyi2017, anderson2015}. Nutrition plays a major role, with ketogenic diets greatly elevating BrAce, and even garlic ingestion can have an impact~\citep{spanel2011, anderson2015}. Consumption of alcohol or certain drugs, such as disulfiram, can increase BrAce 10- to 15-fold~\citep{ruzsanyi2017}. Personal characteristics, including body-mass index (BMI), sex, and exercise habits, as well as disease states, such as diabetes and infections, further contribute to BrAce variation~\citep{anderson2015, turner2006, wang2013, ruzsanyi2017}. Environmental factors, such as exposure to exogenous acetone or even ambient temperature, can affect BrAce levels, too~\citep{fujino1992, anderson2015}. All these factors can potentially confound BrAce-based measurement of fat oxidation~\citep{smith2017}.

Consequently, it seems reasonable to explore exhaled breath for other VOCs that more specifically reflect fat oxidation. Although more than a thousand VOCs have been found in human breath~\citep{drabinska2021}, only few studies have examined the relationship between fat oxidation and compounds other than acetone. Several studies found variations in breath isoprene levels associated with physical activity~\citep{king2009, heaney2022, bell2025}. Yet, recent discoveries implicate isoprene to predominantly originate from muscular lipolytic cholesterol metabolism~\citep{sukul2023}, thereby putting to rest the long-held erroneous belief of its cholesterol biosynthesis via the mevalonate pathway in the liver~\citep{mochalski2024} and negating its utility as a fat-oxidation marker.

A further challenge in monitoring fat oxidation during exercise relates to the presence of ketone bodies, such as BOHB and BrAce. Multiple studies have shown that ketone levels change very little during exercise, with substantial changes occurring only hours later, during the post-exercise recovery phase~\citep{nagamine2022, weber2021, guentner2017, lee2024, kim2020}. Clearly, having to wait so long after training for ketosis to be reliably determined is very inconvenient and offers no benefit during exercise.

To address these challenges, the present study makes two contributions. First, we applied a data-driven, untargeted analytical approach to screen for VOCs that could be potential markers of fat oxidation besides acetone. Second, we assessed whether post-exercise changes in blood BOHB reference measurements could already be predicted from breath measurements during exercise.

\section{Method}

\subsection{Ethics and cohort}

\begin{table}[t]
\centering
\begin{tabular}{lr}
\toprule
Participants & 19 \\
Sex {[}f/m{]} & 6/13 \\
Age {[}years{]} & 30.2 $\pm$ 7.2 \\
BMI {[}kg/m\textsuperscript{2}{]} & 23.6 $\pm$ 2.7 \\
Height {[}cm{]} & 175.5 $\pm$ 8.2 \\
Weight {[}kg{]} & 73.3 $\pm$ 13.1 \\
\bottomrule
\end{tabular}
\caption{Cohort demographics and metadata. Values are presented as mean $\pm$ standard deviation. BMI = body mass index.}\tabularnewline
\label{tab:cohort}
\end{table}

This work was undertaken in accordance with the Declaration of Helsinki and was approved by the Ethics Committee of Friedrich-Alexander Universität Erlangen-Nürnberg (Ethics No. 24-315-S). Informed written consent was obtained from each participant.

A total of 20 participants were recruited for this exploratory study. The data from one participant was later excluded due to corrupted data, such that only the remaining 19 participants were considered in the data evaluation (see Table~\ref{tab:cohort}).

All participants were adults aged between 18 and 50, and self-reported as being neurologically and psychologically healthy. None of the participants were pregnant, breastfeeding or taking medicine for diabetes mellitus, high blood pressure, dyslipidemia, or obesity. All participants were in a physical state to carry out the study.

Participants were instructed to refrain from alcohol consumption, smoking, and intense physical activity for 24 h preceding the experiment. As an optional preparation, they were advised to consume a low-carbohydrate meal on the evening before testing to maximize fat burning during exercise; however, fasting on the day of the experiment was not required. Throughout each experimental session, participants were required to abstain from food and consume only water. No additional dietary restrictions (e.g., standardized meal plans) were imposed to reflect real-life scenarios, and information regarding recent food and beverage intake was collected using participant questionnaires.

\subsection{Experimental design}

\begin{figure*}[t]
\centering
\includegraphics[width=\textwidth]{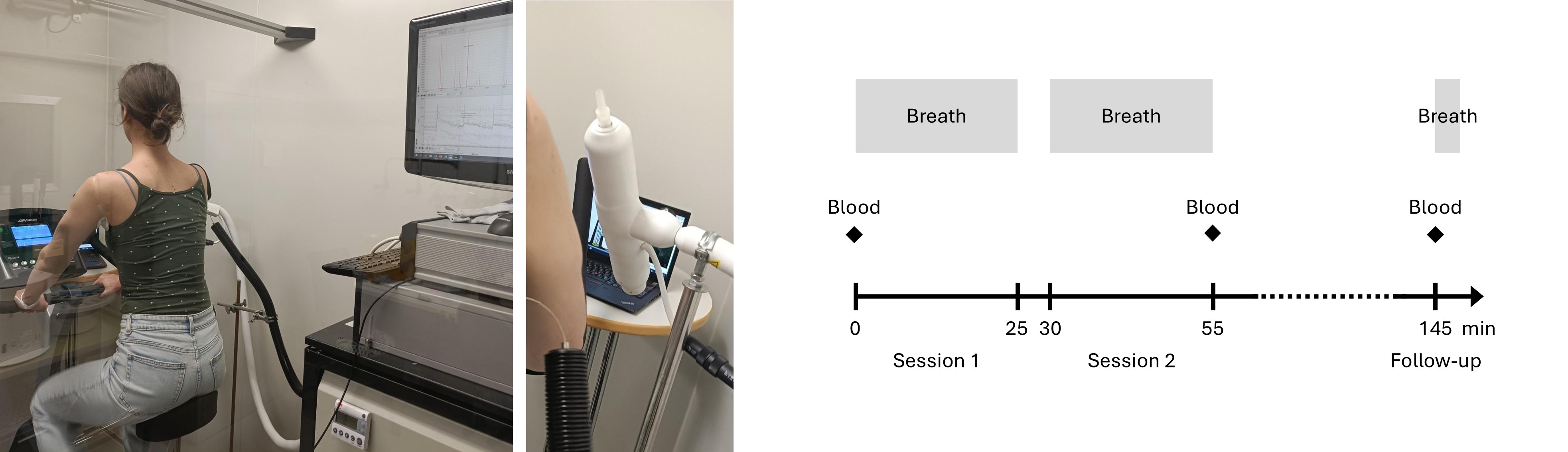}
\caption{Experimental protocol. Left: Participants completed two 25-min cycling sessions on a cycle ergometer, interspersed by a 5-min rest period. Centre: Exhaled breath was sampled via a buffered end-tidal (BET) sampler. A metronome displayed on a notebook helped maintain a constant pedalling rhythm. Right: PTR-TOF-MS measurements of exhaled breath were recorded throughout both cycling sessions and at follow-up (90 min post-exercise). Blood BOHB concentrations were determined immediately before and after the cycling sessions and at follow-up.}
\label{fig:experimental_protocol}
\end{figure*}

All participants completed a standardized experimental protocol on an upright cycle ergometer (C1 Upright Lifecycle, Life Fitness Europe GmbH, Unterschleißheim, Germany). The protocol was designed to be similar to those used in related studies~\citep{guentner2017, koenigstein2020}. The experimental sequence is illustrated in Figure~\ref{fig:experimental_protocol}.

Participants arrived at the laboratory 15 min prior to the start of the experiment to complete preparatory steps. During this period, the study protocol was explained, written informed consent forms and data protection declarations were signed, and a questionnaire was completed. The cycle ergometer was adjusted to match the individual height of the participant.

Each participant completed two 25-min cycling sessions, separated by a 5 min rest period. At the beginning and during the cycling sessions, various measurements were taken, as described below. A metronome set to 132 beats per minute (corresponding to 66 rounds per minute) was provided to maintain a constant pedalling rhythm. Exercise intensity was adjusted individually to achieve a workload perceived as moderately exhausting and sustainable for the duration of the protocol. Resistance was modified during the sessions if participants showed signs of overexertion. After the second cycling session, participants rested for 90 min under fasting conditions (only water allowed) before follow-up measurements were taken.

\subsection{PTR-TOF-MS measurements}

Volatiles in exhaled breath were analysed using a proton-transfer-reaction time-of-flight mass spectrometer (PTR-TOF-MS) instrument (PTR-TOF 8000, IONICON Analytik GmbH, Innsbruck, Austria). PTR-TOF-MS uses hydronium ions to ionize molecules via proton transfer reactions in a drift tube~\citep{ellis2013}. The resulting ions are then separated and detected using a time-of-flight mass spectrometer. PTR-TOF-MS enables real-time measurements with high temporal and mass resolutions and high sensitivity, making it ideal for non-targeted, time-resolved analyses. 

Measurements were performed continuously during both cycling sessions. While exercising, participants were instructed to perform two exhalations within 1 min every 2 min. After the 90-min rest period, measurement was resumed for approximately 5 min in the same manner as during the cycling sessions to obtain follow-up values.

Breath was sampled using a buffered end-tidal (BET) breath sampler (IONICON Analytik GmbH, Innsbruck, Austria) connected to the PTR-TOF-MS instrument~\citep{herbig2008}. Disposable mouthpieces (EnviteC Wismar GmbH, Wismar, Germany) were used for each new participant to maintain hygiene and avoid carry-over from the previous participant.

\subsection{BOHB measurements}

Blood-based BOHB concentrations were used as reference values for the level of fat oxidation, similar to related studies~\citep{suntrup2020, delorbe2022, guentner2017, laffel1999}. BOHB concentrations are less susceptible to interference from other substances in food or pharmaceuticals than BrAce~\citep{anderson2015, laffel1999}. Moreover, BOHB does not rely on the diffusion of molecules through the lungs, making it a more direct measure of fat oxidation~\citep{kim2020}.

BOHB concentrations were determined by taking capillary blood samples from the fingertips of participants using FORA Sterile Lancets (SMART OTC GmbH, Mannheim, Germany). These were then applied to FORA β-Ketone test strips (SMART OTC GmbH, Mannheim, Germany), which were inserted into a FORA 6 Duo Multi-Functional Monitoring System (SMART OTC GmbH, Mannheim, Germany). The device displayed the BOHB level (mmol/L) after about 10 s within a detection range of 0.1 to 8.0 mmol/L. BOHB measurements were taken at three time points for each participant: immediately before the first cycling session, directly after the second cycling session, and at follow-up.

\subsection{Data pre-processing}

The raw PTR-TOF-MS measurement data were pre-processed and harmonized. Raw PTR-TOF-MS data are organized as two-dimensional matrices, with columns representing recording cycles (each lasting 1 s) and rows representing bins in the recorded TOF spectrum. The matrix values reflect ion counts detected at each cycle and TOF bin. TOF bins are monotonically related to mass-to-charge ratios (\emph{m/z}) and thus allow differentiation between various ion species.

\begin{figure*}[t]
\centering
\includegraphics[width=\textwidth]{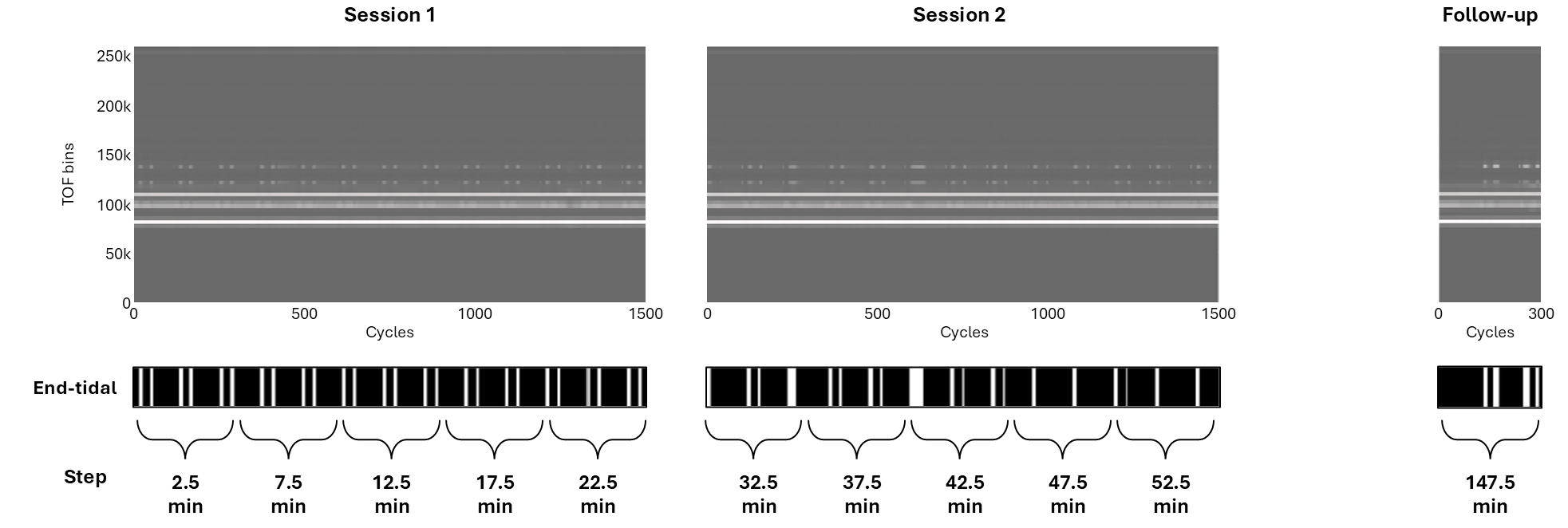}
\caption{Data pre-processing. The raw PTR-TOF-MS measurements of every participant were condensed into eleven 5-minute steps. For each step, a TOF spectrum was computed by masking the cycles belonging to end-tidal phases and averaging the respective TOF bin values.}
\label{fig:data_preprocessing}
\end{figure*}

Within the PTR-TOF-MS measurements, only the end-tidal phase of each exhalation was considered relevant, since this represents gas from deeper in the lungs that contains the highest systemic metabolite concentrations~\citep{lee2024}. To automatically detect the end-tidal plateaus, K-means clustering-based automated thresholding~\citep{liu2009} was applied to produce masks marking the associated measurement cycles (Figure~\ref{fig:data_preprocessing}). Visual inspection confirmed the accuracy of all generated masks.

The PTR-TOF-MS data were then transformed into a uniform data structure that enables comparison of progressions across participants. The raw data were condensed into sequences of 5-min steps. Each of the two cycling sessions was divided into five steps, and the follow-up measurement into one step, resulting in 11 steps per participant. The steps were labelled according to their midpoint in minutes (Figure~\ref{fig:data_preprocessing}). First, the end-tidal masks were applied to select the relevant cycles. Then, the TOF bin values of all selected cycles were averaged to generate one TOF spectrum per step. A duration of 5 min per step was selected to ensure that at least four end-tidal phases were captured in each step. The final harmonized data formed a three-dimensional matrix in the format of 19 participants $\times$ 11 steps $\times$ 260,000 TOF bins, with the values reflecting the ion counts recorded in each step and TOF bin.

\subsection{Signal detection}

In PTR-TOF-MS spectra, peak-shaped signals indicate the presence of certain ion species, with the TOF bin at the signal maximum roughly indicating their mass-to-charge ratio (\emph{m/z})~\citep{ellis2013}. We developed a simple algorithm to detect relevant peak signals in an untargeted manner, making only minimal assumptions about peak shapes or scales.

In the first stage, signals were detected separately for each step (of each participant) and mapped to \emph{m/z} values. Each step's 260,000 TOF bins spectrum was smoothed using a 1D Gaussian convolution filter (σ~=~10) to reduce noise~\citep{burger2009}. Signals were then detected at TOF bins whose immediate neighbours both had smaller values. When multiple adjacent TOF bins shared the same maximum value, the central bin was chosen. Detected TOF bins were mapped to \emph{m/z} values using the common two-constant calibration function $\mathit{m/z} = ((\mathit{TOF\_bin} - \mathit{p2}) / \mathit{p1})^2$~\citep{mueller2013}. To account for spectral drift during the measurements, the calibration function was re-fitted for each step based on the signals of the hydronium primary ion (\emph{m/z}~21.022) and the iodobenzene ion signal (\emph{m/z}~203.943)~\citep{mueller2013} arising from the permanent mass scale calibration (PerMasSCal) feature of the instrument.

In the second stage of the algorithm, a set of relevant signals was identified by matching signals across steps. All signals detected in all steps were clustered using the density-based DBSCAN algorithm ($\epsilon~=~0.015~\mathit{m/z}$)~\citep{schubert2017}. Signals were considered matching if they were part of the same cluster. Clusters containing signals detected in at least ten different participants were deemed relevant. Each relevant cluster was represented by the mean \emph{m/z} value of the signals it contained, resulting in a set of \emph{m/z} values of relevant signals.

\subsection{Signal quantification}

The strength of a PTR-TOF-MS signal is (ideally) proportional to the abundance of the corresponding ion species. We opted to quantify signal strength based on signal intensity (peak amplitude) rather than signal area. Isolated signals in PTR-TOF MS spectra generally have a consistent shape~\citep{cappellin2011}, meaning that their intensity and area are proportional, which was confirmed by our dataset. Absolute area values were not required for our investigation, as we focused solely on signal ratios. Calculating signal intensity works without any assumptions about baselines and integration limits required for area calculation~\citep{cappellin2011}. We did not attempt to separate overlapping signals, as we expected most meaningful signals to represent only a single ion species, and because a generally functional signal separation, needed for non-targeted analysis, would have been difficult to implement~\citep{cappellin2011}. Random verification confirmed that the signals selected for further scrutiny did not overlap.

Signal intensities were quantified in each step individually, based on the smoothed PTR-TOF-MS spectra described before. For each relevant signal (representing a cluster mean as described above), the nearest signal detected in that step was identified within a tolerance of $\pm$0.015~\emph{m/z}. If a suitable signal was found, its maximum ion count was taken as the signal intensity. Otherwise, the intensity was considered zero. Finally, the intensity values were normalised by the intensity of the hydronium ion signal from the same step, making them comparable~\citep{ellis2013}.

\subsection{Statistical analysis}

Unless stated otherwise, all references to signal values or BOHB levels in the following sections denote their relative changes from the initial time point (2.5 min). This enables evaluation of the effects of exercise and comparison between participants with different baseline concentrations~\citep{guentner2017, laffel1999}. No dedicated baseline PTR-TOF-MS measurements were acquired immediately before exercise. Previous studies have shown that fat oxidation is negligible during the first few minutes of exercise~\citep{koenigstein2020, guentner2017}. Consequently, we assumed that the intensities of signals associated with fat oxidation at the initial time point were good approximations of the intensities immediately before exercise. Correlations between signal values and BOHB levels were generally evaluated using the Spearman rank correlation coefficient (ρ), which can identify non-linear monotonic relationships~\citep{anderson2015}. Calculation of p-values was made using a non-parametric permutation test with 9999 resamples.

To assess the predictability of follow-up BOHB levels from end-of-exercise signal values, K-nearest neighbour regression models were trained and applied using leave-one-out cross-validation. For each signal and each participant, a distinct model was trained on data from all other participants, using the end-of-exercise signal values as input and the corresponding follow-up BOHB levels as output. These models were then applied to the respective held-out participants to generate independent predictions of follow-up BOHB levels. Regression quality was assessed in terms of the absolute error distributions between the observed and predicted BOHB levels of the held-out participants.

In addition to regression quality, we evaluated whether end-of-exercise BrAce measurements could serve as early indicators for identifying participants who would exhibit substantial post-exercise changes in BOHB. To this end, both observed and predicted BOHB levels were binarized into "marginal" or "substantial" groups by applying a threshold value. This threshold was determined empirically to ensure clear separation between both groups (see Results). Classification performance was then assessed using F1 scores and accuracy, calculated from the binarized BOHB levels.

\section{Results}

\subsection{Untargeted search for volatile markers in breath}

The automated signal detection algorithm identified 773 relevant signals. To identify signals linked with ketone metabolism, we analysed correlations between the relative changes of the relevant signals and BOHB. Since it is known that substantial changes in fat oxidation occur only after exercise, these correlations were assessed at the follow-up time point (147.5~min). Table~\ref{tab:untargeted_search_results} lists the ten signals with the highest absolute correlation values. Only four signals (\emph{m/z}~31.027, 59.046, 59.214 and 60.051) showed strong correlations with BOHB levels (ρ~≥~0.82, p~=~0.0002), while all other signals exhibited much weaker correlations (ρ~≤~0.57, p~≥~0.0126). 

\begin{table}
\centering
\begin{tabular}{lr}
\toprule
\textbf{Signal (\textit{m/z})} & \textbf{Spearman's ρ} \\ \midrule
\textbf{59.214} & \textbf{0.90 (p = 0.0002)} \\
\textbf{31.027} & \textbf{0.88 (p = 0.0002)} \\
\textbf{60.051} & \textbf{0.83 (p = 0.0002)} \\
\textbf{59.046} & \textbf{0.82 (p = 0.0002)} \\
20.025 & 0.57 (p = 0.0126) \\
17.027 & -0.55 (p = 0.0160) \\
67.056 & 0.55 (p = 0.0154) \\
38.034 & 0.47 (p = 0.0460) \\
29.999 & 0.46 (p = 0.0506) \\
7.337 & 0.44 (p = 0.0732) \\
\bottomrule
\end{tabular}
\caption{Untargeted search results. The ten of the 773 relevant signals with the highest absolute Spearman coefficient values at the follow-up time point. Sorted from high to low.}
\label{tab:untargeted_search_results}
\end{table}

\begin{figure}
\centering
\includegraphics[width=\columnwidth]{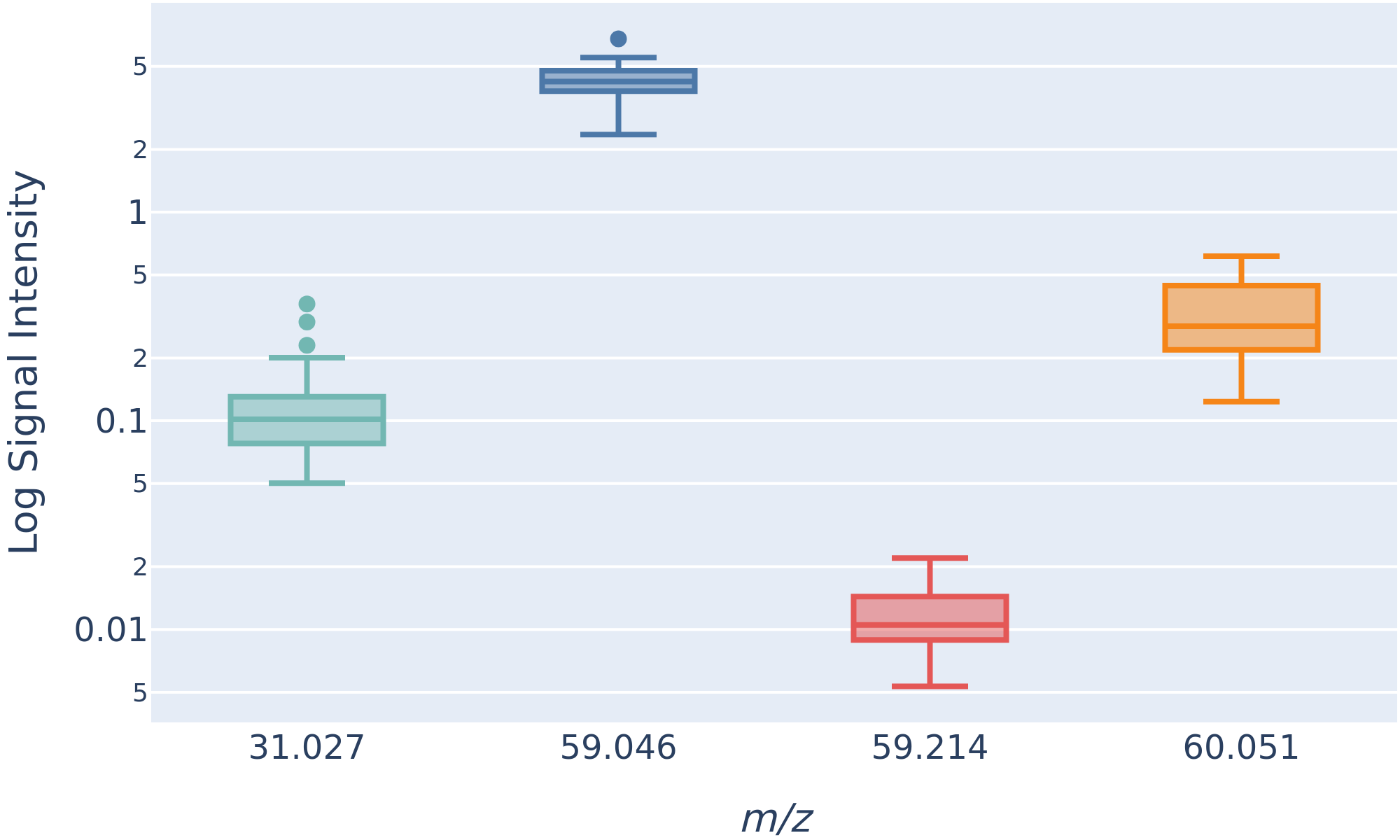}
\caption{Distributions of the logarithmic intensities of the four significant signals at the follow-up measurement. Data comprise the mean end-tidal plateau signal intensities of four exhalations at follow-up for each participant.}
\label{fig:raw_value_distribution}
\end{figure}

The four \emph{m/z} with the strongest correlations to BOHB also had high intercorrelations, as depicted in the Spearman correlation coefficient matrix in Table~\ref{tab:signal_correlation}, indicating a common origin. Figure~\ref{fig:raw_value_distribution} visualizes distributions of the raw logarithmic intensity values (not their relative changes) of these four signals. Each data point represents one participant. Evidently, the signals occur at different magnitudes, with \emph{m/z}~59.046 being by far the most abundant.

\begin{table}
\centering
\begin{tabular}{l|rrr}
\toprule
\textbf{Signal (\textit{m/z})} & \textbf{59.046} & \textbf{59.214} & \textbf{60.051} \\ \midrule
\textbf{31.027} & 0.97 & 0.99 & 0.91 \\
\textbf{59.046} &      & 0.98 & 0.89 \\
\textbf{59.214} &      &      & 0.91 \\ 
\bottomrule
\end{tabular}
\caption{Spearman correlation coefficients between the four signals at the follow-up time point.}
\label{tab:signal_correlation}
\end{table}

With this information and based on visual interpretation of the peak shapes, the signals at \emph{m/z}~59.046 and \emph{m/z}~60.051 could be attributed to protonated acetone (\emph{m/z}~59.049) and its \ce{^{13}C}-isotopologue (\emph{m/z}~60.052), respectively. However, a deviation in the ratio of their abundances ($\sim$7.5 \% observed vs 3.3 \% expected) means that a contribution to the signal at \emph{m/z}~60.051 from an additional compound of unknown identity cannot be ruled out. Of the remaining two signals, \emph{m/z}~31.027 likely represents \ce{CH2OH^+} (\emph{m/z}~31.018), which could be attributable to formaldehyde and/or a negligible fragment of acetone: the former is inherently difficult to detect in PTR-MS due its similar proton affinity to water and humidity modulation~\citep{inomata2008, hansel1997}; the latter has only previously been reported in ion trap-based PTR-MS studies~\citep{prazeller2003, warneke2005}, but nevertheless could be present in conventional PTR-TOF-MS in very low abundances, as observed here (< 3 \%). Finally, the signal at \emph{m/z}~59.214 could not be clearly attributed to any known compound, but the very strong correlation with acetone and its very low intensity suggest that this is an instrumental artefact of the main acetone signal. Overall, the slight deviations from the reference mass values are likely due to tolerances in automatic calibration and signal detection.

Visual scrutiny of the data allowed the six weakly correlated signals to be attributed to signal artefacts: \emph{m/z} 7.337 and \emph{m/z} 67.056 represent low-level signals of unknown origin that could not be linked to any specific ions; the remaining \emph{m/z} listed are common signals in PTR-TOF-MS datasets associated with ion species generated as minor products in the hollow cathode ion source of the instrument. Specifically, \emph{m/z} 17.027 is likely charged ammonia (\ce{NH3^+}, \emph{m/z} 17.026; commonly mainly detected as a protonated ion, i.e., \ce{NH4^+}, \emph{m/z} 18.026); \emph{m/z} 20.025 is a signal peak artefact in the tail of the highly saturated protonated water peak (\emph{m/z} 19.018); and \emph{m/z} 29.999 is attributable to \ce{NO^+} (\emph{m/z} 29.997). 

Protonated isoprene and one of its fragments were detected at signals \emph{m/z}~69.070 and \emph{m/z}~41.040, but both showed only weak correlations with BOHB levels in end-tidal measurements (each~ρ~=~0.28, p~≥~0.2444).

\subsection{Prediction of post-exercise fat oxidation}

Our second objective was to investigate the extent to which fat oxidation after exercise can be predicted during exercise. For this, we focused on the signals of protonated acetone and its \ce{^{13}C}-isotopologue (detected at \emph{m/z}~59.046 and \emph{m/z}~60.051) that showed clear correlations with BOHB and represented intact analyte ions.

\begin{figure*}
\centering
\includegraphics[width=\textwidth]{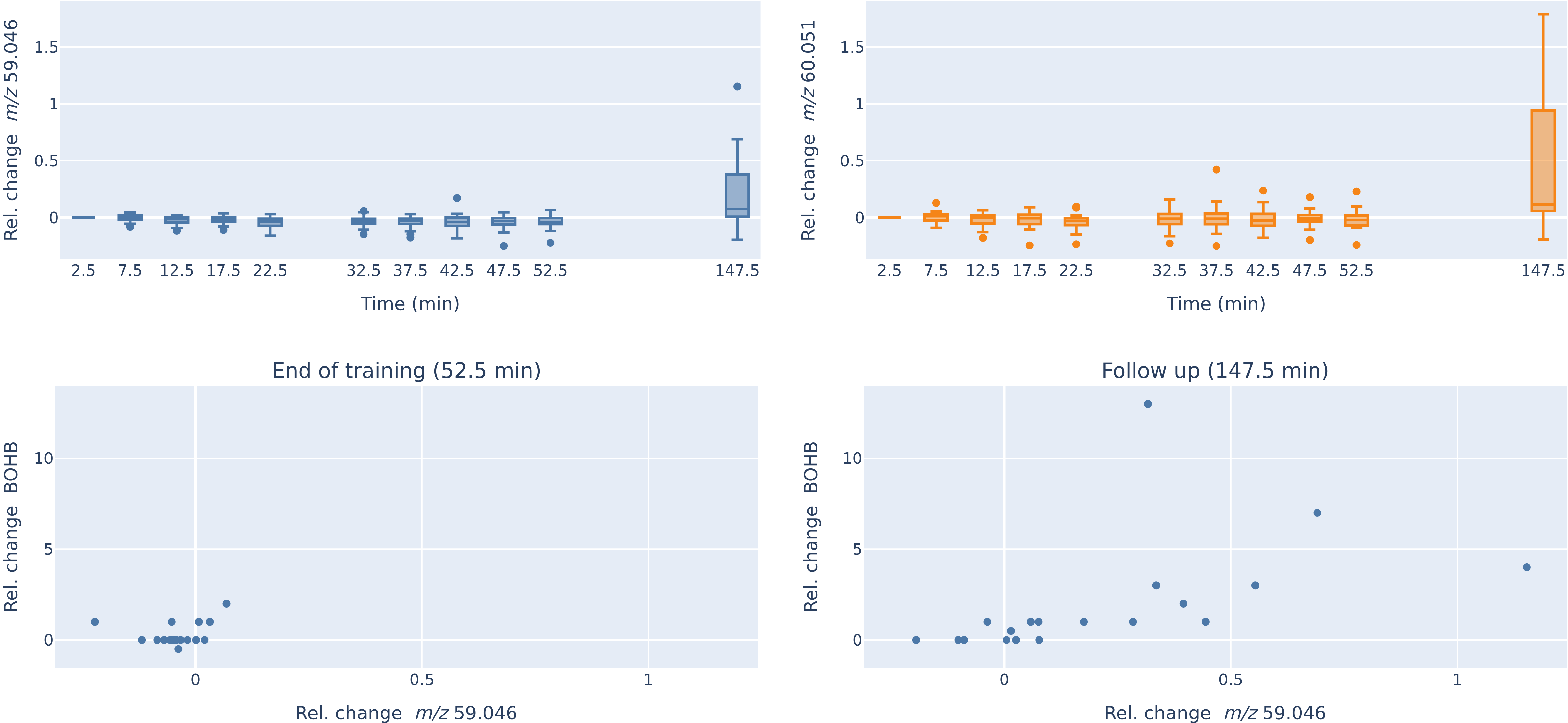}
\caption{Signal changes over time. Top: Distributions of relative changes in signal intensity at \emph{m/z}~59.046 (left) and \emph{m/z}~60.051 (right) compared to the start of exercise. The signals represent protonated acetone and its \ce{^{13}C}-isotopologue, respectively. Bottom: Correlation plots of the relative changes in \emph{m/z}~59.046 and BOHB at the end of training (52.5~min; left) and the follow-up time point (147.5~min; right).}
\label{fig:time_course}
\end{figure*}

The top row of Figure~\ref{fig:time_course} visualizes the distributions of the changes of the two acetone signals across all participants over time relative to at the start of exercise. During exercise, these values changed only slightly, generally tending to decrease. However, at follow-up, they increased sharply. The distributions at follow-up were positively skewed, meaning most participants showed small increases, while a few had much larger ones.

Figure~\ref{fig:time_course}, bottom row, plots the relative changes in protonated acetone compared to BOHB at the end of exercise (52.5~min) and follow-up (147.5~min). BOHB exhibits a similar trend to the two acetone signals. At the end of exercise, BOHB values remain largely unchanged, with most participants showing no variation at all. A notable increase in BOHB values only occurs at the follow-up time point.

\begin{figure*}
\centering
\includegraphics[width=\textwidth]{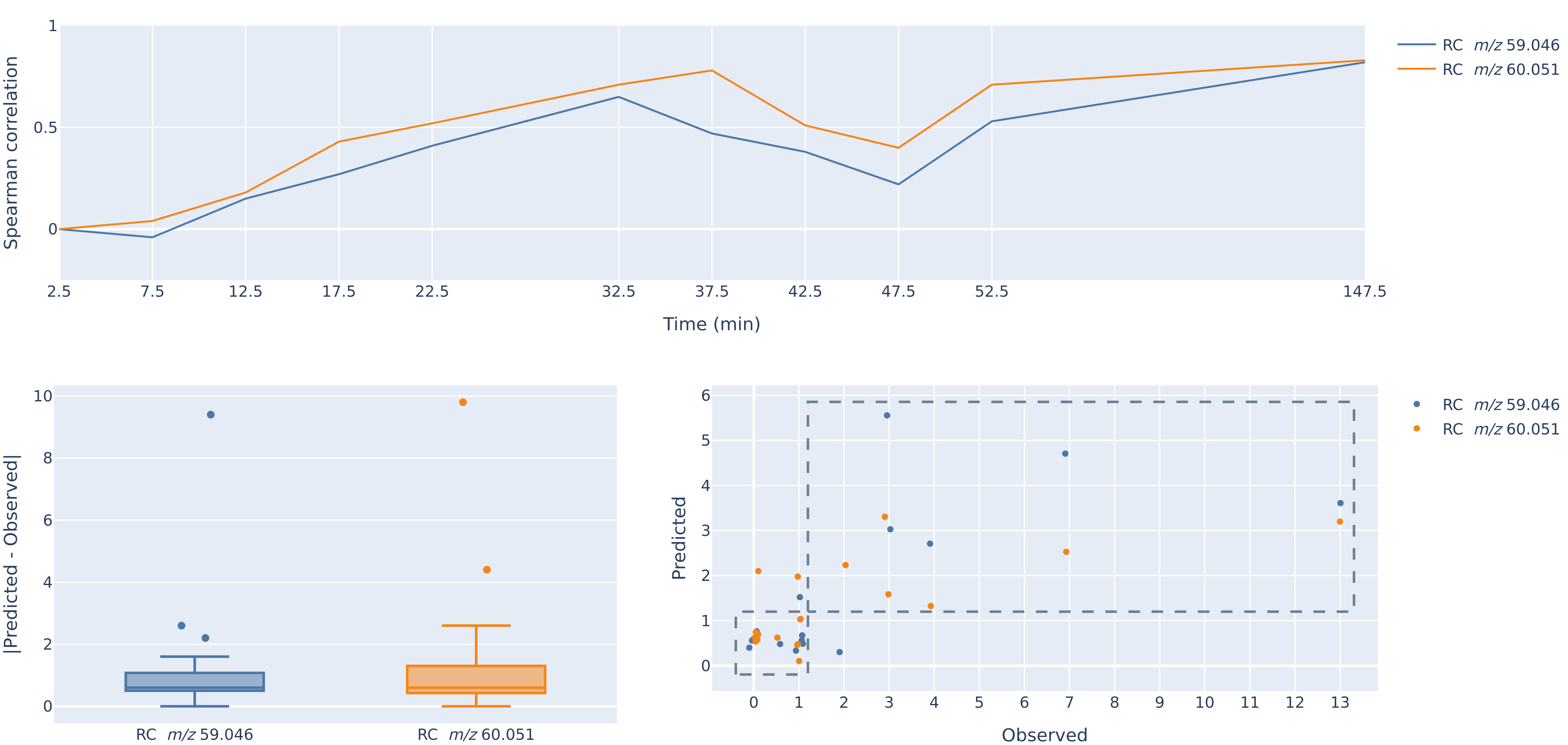}
\caption{Post-exercise predictability. Top: Spearman correlations between relative changes in signal intensity of the two acetone signals (\emph{m/z}~59.046, \emph{m/z}~60.051) at different time points and relative changes in BOHB at follow-up. Bottom: Predictability of relative changes in BOHB at follow-up based on relative changes in the acetone signals at the end of exercise: absolute error distributions (left) and scatterplot of observed versus predicted values (right).}
\label{fig:post_exercise_predictability}
\end{figure*}

Figure~\ref{fig:post_exercise_predictability}, top row, shows how the acetone signals at different time points correlated with BOHB at follow-up. Until the beginning of the second cycling session, the correlations increased continuously. However, in the middle of the second session, there was a notable decline in correlation, which party recovered by the end of the session.

Figure~\ref{fig:post_exercise_predictability}, bottom row, illustrates the predictability of BOHB at follow-up based on the acetone signals at the end of exercise. The left chart shows the absolute error distributions for each signal, representing the absolute differences between observed and predicted relative changes of BOHB levels at follow-up. No signal exhibited  clearly superior predictive power.

The scatterplot on the right compares observed and predicted post-exercise BOHB values. Dense clustering in the lower left corner indicates that approximately two-thirds of participants (13/19) exhibited only marginal relative changes in BOHB levels, whereas the remaining six participants showed substantial changes. These two groups are separated by a threshold of~$\sim$1.2. The dashed boxes indicate participants correctly predicted to belong to either group. Using end-of-exercise values of protonated acetone, participants with substantial post-exercise changes in BOHB (≥1.2) could be correctly identified with an F1~score of~0.83 (accuracy:~0.89). Classification based on the corresponding \ce{^{13}C}-isotopologue values yielded comparable performance, with an F1~score of approximately~0.86 (accuracy:~0.89).

\section{Discussion}

In this study, we took an untargeted approach to identify volatile markers of exercise-induced fat oxidation. Using a fully data-driven analysis method, we detected 773~relevant signals in PTR-TOF-MS measurements of end-tidal breath. We considered post-exercise changes in BOHB as the best indicator of the extent of fat oxidation achieved through exercise. Signals associated with fat oxidation were therefore expected to show strong correlations with BOHB at the follow-up time point. Among the 773 relevant signals, only four signals exhibited very strong correlations with BOHB, while all remaining signals showed much weaker associations, which were considered inconclusive due to the small sample size.

The four strongly correlated signals could all be linked to acetone. Two signals were attributed to protonated acetone and its \ce{^{13}C}-isotopologue, respectively, while the others likely represent a minor fragment and a signal artefact, respectively. Our results confirm previous studies that found a strong correlation between BrAce and BOHB. Also, the temporal concentration profiles of these signals (Figure~\ref{fig:time_course}), including a sharp increase after exercise, are consistent with prior studies~\citep{nagamine2022, weber2021, guentner2017, lee2024, kim2020}. Apart from acetone, our study did not provide any evidence for meaningful breath-based markers of fat oxidation. Isoprene, which has been linked to physical activity in some previous studies, did not correlate with BOHB to any meaningful degree.

To our knowledge, this is the first study to investigate whether post-exercise changes in BOHB can be predicted from measurements taken during exercise. In our cohort, approximately one third of participants showed substantial post-exercise increases in BOHB, whereas the remaining participants showed only marginal changes (Figure~\ref{fig:post_exercise_predictability}). A similar pattern was reported in other studies~\citep{delorbe2022}. By the end of both cycling sessions, it was possible to accurately predict whether participants would have substantial changes in BOHB 90 min later based on their current changes in BrAce. This suggest that it could be possible to make basic statements about the expected fat oxidation already during exercising and solely based on BrAce measurements.

However, our results also suggest that the predictability of post-exercise fat oxidation may vary over the course of exercise and may be influenced by intermittent breaks. While the correlation between BrAce during exercise and follow-up BOHB rose steadily throughout the first cycling session, it showed a temporary decline during the second session, for unknown reasons.

\subsection{Limitations}

We verified the accuracy of our signal detection algorithm by randomly inspecting detected peaks and confirming the identification of compounds commonly found in breath. The consistency of our results with previous studies further supports the validity of the algorithm. While we did not attempt to separate overlapping signals, random inspection rarely revealed peaks so closely spaced that their detection would be impaired. A quantitative accuracy evaluation would have required comprehensive reference annotations of signals, which was beyond the scope of this work. Therefore, we cannot rule out the possibility that some signals were missed.

To ensure comparability between measurements, we only considered a uniform exercise protocol consisting of two cycling sessions of equal duration. Hence, our results do not allow any conclusions to be drawn about the predictability of fat oxidation in other exercise scenarios. It is important to note that the reported time courses of correlations between signal values and BOHB levels only reflect the scenario of the described experimental protocol in which both cycling sessions were completed in full.

\subsection{Future directions}

Future research should investigate more thoroughly whether BrAce measurements during exercise can serve as reliable predictors of post-exercise fat oxidation. Studies should utilize diverse exercise protocols to assess how training duration, interruptions and personal, nutritional, physiological and environmental factors affect predictive accuracy. To better characterize post-exercise changes, these studies should include measurements at multiple follow-up time points. Achieving adequate statistical power will require considerably larger sample sizes and more diverse cohorts. Ideally, studies should also include measurements of acetone made using low-cost, portable gas sensors that can be integrated into end-user devices~\citep{bastide2023, lee2024, delorbe2022, weber2021, kim2020, guentner2017}, to assess whether these devices can resolve acetone concentrations with sufficient sensitivity and specificity for reliable predictions.

If the predictability of post-exercise fat oxidation proves feasible more generally, future BrAce monitoring devices could provide users with real-time estimates of anticipated fat oxidation by incorporating suitable prediction models. These systems would be much more practical than the existing devices because users would no longer have to wait several hours after training to obtain meaningful results. Moreover, immediate feedback could boost motivation to train.

\section{Conclusion}

Untargeted analysis of end-tidal breath by PTR-OF-MS did not reveal any novel markers of exercise-induced fat oxidation. Nevertheless, our results corroborated earlier research showing a strong correlation between BrAce and blood BOHB, with substantial elevations in these ketone bodies observed only during the post-exercise recovery phase. To our knowledge, this is the first study examining whether breath measurements taken during exercise can predict subsequent post-exercise changes in BOHB. Our findings suggest that basic predictions of eventual fat oxidation may be feasible during training, based solely on BrAce measurements. If subsequent research validates these findings more broadly, this could enable the development of innovative BrAce monitoring devices that provide users with real-time projections of anticipated fat oxidation during physical activity.

\section*{Acknowledgements}

This research received funding from the Fraunhofer Society for the BurnOn! project as part of the Discover funding programme.

The authors would like to thank all study participants for undertaking the effort of completing the cycling sessions and measurements.

\section*{Author contributions}

Conceptualization: AH, JB, and YLP. Data Curation: AH, JBS, and YLP. Formal Analysis: AH and YLP. Investigation: AH, JB and YLP. Methodology: AH, JB, and YLP. Software: AH and JPR. Supervision: AH, JB, and YLP. Project Administration: AH, JB, and YLP. Funding Acquisition: AH, JB, and YLP. Writing---Original Draft Preparation: AH. Writing---Review and Editing: AH, JBS, JPR, JB, and YLP.

\section*{Data availability}

The data cannot be made publicly available upon publication because they contain sensitive personal information. The data that support the findings of this study are available upon reasonable request from the authors.

\section*{Conflict of interest}

The authors declare that there is no conflict of interest.

\section*{Ethical statement}

This work was undertaken in accordance with the Declaration of Helsinki and was approved by the Ethics Committee of Friedrich-Alexander Universität Erlangen-Nürnberg (Ethics No. 24-315-S). Informed written consent was obtained from each participant.

\bibliographystyle{vancouver}
\bibliography{references}

@article{achten2004,
  title = {Optimizing fat oxidation through exercise and diet},
  volume = {20},
  ISSN = {0899-9007},
  url = {http://dx.doi.org/10.1016/j.nut.2004.04.005},
  DOI = {10.1016/j.nut.2004.04.005},
  number = {7–8},
  journal = {Nutrition},
  publisher = {Elsevier BV},
  author = {Achten,  Juul and Jeukendrup,  Asker E},
  year = {2004},
  month = jul,
  pages = {716–727}
}

@inbook{abegg2020,
  title = {Lifestyle applications},
  ISBN = {9780128199671},
  url = {http://dx.doi.org/10.1016/B978-0-12-819967-1.00023-2},
  DOI = {10.1016/b978-0-12-819967-1.00023-2},
  booktitle = {Breathborne Biomarkers and the Human Volatilome},
  publisher = {Elsevier},
  author = {Abegg,  Sebastian and G\"{u}ntner,  Andreas T.},
  year = {2020},
  pages = {363–375},
  chapter = 23
}

@article{amarogahete2019,
  title = {Assessment of maximal fat oxidation during exercise: A systematic review},
  volume = {29},
  ISSN = {1600-0838},
  url = {http://dx.doi.org/10.1111/sms.13424},
  DOI = {10.1111/sms.13424},
  number = {7},
  journal = {Scandinavian Journal of Medicine \& Science in Sports},
  publisher = {Wiley},
  author = {Amaro‐Gahete,  Francisco J. and Sanchez‐Delgado,  Guillermo and Jurado‐Fasoli,  Lucas and De‐la‐O,  Alejandro and Castillo,  Manuel J. and Helge,  Jørn W. and Ruiz,  Jonatan R.},
  year = {2019},
  month = apr,
  pages = {910–921}
}

@article{anderson2015,
  title = {Measuring breath acetone for monitoring fat loss: Review},
  volume = {23},
  ISSN = {1930-739X},
  url = {http://dx.doi.org/10.1002/oby.21242},
  DOI = {10.1002/oby.21242},
  number = {12},
  journal = {Obesity},
  publisher = {Wiley},
  author = {Anderson,  Joseph C.},
  year = {2015},
  month = nov,
  pages = {2327–2334}
}

@article{bastide2023,
  title = {Handheld device quantifies breath acetone for real-life metabolic health monitoring},
  volume = {2},
  ISSN = {2635-0998},
  url = {http://dx.doi.org/10.1039/d3sd00079f},
  DOI = {10.1039/d3sd00079f},
  number = {4},
  journal = {Sensors \& Diagnostics},
  publisher = {Royal Society of Chemistry (RSC)},
  author = {Bastide,  Grégoire M. G. B. H. and Remund,  Anna L. and Oosthuizen,  Dina N. and Derron,  Nina and Gerber,  Philipp A. and Weber,  Ines C.},
  year = {2023},
  pages = {918–928}
}

@article{bell2025,
  title = {Reliability of Exhaled Acetone and Isoprene in Healthy Active Adult Responses to Sub-maximal Treadmill Exercise: A Pilot Study},
  ISSN = {2662-1371},
  url = {http://dx.doi.org/10.1007/s42978-025-00327-x},
  DOI = {10.1007/s42978-025-00327-x},
  journal = {Journal of Science in Sport and Exercise},
  publisher = {Springer Science and Business Media LLC},
  author = {Bell,  Leo R. and Myers,  Mark A. and Harvey,  Jack T. and Hennessy,  Declan and Worn,  Ryan L. and Wallen,  Matthew P. and Davis,  Greg and O’Brien,  Brendan J.},
  year = {2025},
  month = apr 
}

@book{burger2009,
  title = {Principles of Digital Image Processing: Fundamental Techniques},
  ISBN = {9781848001916},
  ISSN = {1863-7310},
  url = {http://dx.doi.org/10.1007/978-1-84800-191-6},
  DOI = {10.1007/978-1-84800-191-6},
  journal = {Undergraduate Topics in Computer Science},
  publisher = {Springer London},
  author = {Burger,  Wilhelm and Burge,  Mark James},
  year = {2009}
}

@article{cappellin2011,
  title = {On data analysis in PTR-TOF-MS: From raw spectra to data mining},
  volume = {155},
  ISSN = {0925-4005},
  url = {http://dx.doi.org/10.1016/j.snb.2010.11.044},
  DOI = {10.1016/j.snb.2010.11.044},
  number = {1},
  journal = {Sensors and Actuators B: Chemical},
  publisher = {Elsevier BV},
  author = {Cappellin,  Luca and Biasioli,  Franco and Granitto,  Pablo M. and Schuhfried,  Erna and Soukoulis,  Christos and Costa,  Fabrizio and M\"{a}rk,  Tilmann D. and Gasperi,  Flavia},
  year = {2011},
  month = jul,
  pages = {183–190}
}

@article{delorbe2022,
  title = {Breath analyzer for personalized monitoring of exercise-induced metabolic fat burning},
  volume = {369},
  ISSN = {0925-4005},
  url = {http://dx.doi.org/10.1016/j.snb.2022.132192},
  DOI = {10.1016/j.snb.2022.132192},
  journal = {Sensors and Actuators B: Chemical},
  publisher = {Elsevier BV},
  author = {Del Orbe,  Dionisio V. and Park,  Hyung Ju and Kwack,  Myung-Joon and Lee,  Hyung-Kun and Kim,  Do Yeob and Lim,  Jung Gweon and Park,  Inkyu and Sohn,  Minji and Lim,  Soo and Lee,  Dae-Sik},
  year = {2022},
  month = oct,
  pages = {132192}
}

@article{drabinska2021,
  title = {A literature survey of all volatiles from healthy human breath and bodily fluids: the human volatilome},
  volume = {15},
  ISSN = {1752-7163},
  url = {http://dx.doi.org/10.1088/1752-7163/abf1d0},
  DOI = {10.1088/1752-7163/abf1d0},
  number = {3},
  journal = {Journal of Breath Research},
  publisher = {IOP Publishing},
  author = {Drabińska,  Natalia and Flynn,  Cheryl and Ratcliffe,  Norman and Belluomo,  Ilaria and Myridakis,  Antonis and Gould,  Oliver and Fois,  Matteo and Smart,  Amy and Devine,  Terry and Costello,  Ben De Lacy},
  year = {2021},
  month = apr,
  pages = {034001}
}

@book{ellis2013,
  title = {Proton Transfer Reaction Mass Spectrometry: Principles and Applications},
  ISBN = {9781118682883},
  url = {http://dx.doi.org/10.1002/9781118682883},
  DOI = {10.1002/9781118682883},
  publisher = {Wiley},
  author = {Ellis,  Andrew M. and Mayhew,  Christopher A.},
  year = {2013},
  month = dec 
}

@article{fujino1992,
  title = {Biological monitoring of workers exposed to acetone in acetate fibre plants.},
  volume = {49},
  ISSN = {1351-0711},
  url = {http://dx.doi.org/10.1136/oem.49.9.654},
  DOI = {10.1136/oem.49.9.654},
  number = {9},
  journal = {Occupational and Environmental Medicine},
  publisher = {BMJ},
  author = {Fujino,  A and Satoh,  T and Takebayashi,  T and Nakashima,  H and Sakurai,  H and Higashi,  T and Matumura,  H and Minaguchi,  H and Kawai,  T},
  year = {1992},
  month = sep,
  pages = {654–657}
}

@article{guentner2017,
  title = {Noninvasive Body Fat Burn Monitoring from Exhaled Acetone with Si-doped WO3-sensing Nanoparticles},
  volume = {89},
  ISSN = {1520-6882},
  url = {http://dx.doi.org/10.1021/acs.analchem.7b02843},
  DOI = {10.1021/acs.analchem.7b02843},
  number = {19},
  journal = {Analytical Chemistry},
  publisher = {American Chemical Society (ACS)},
  author = {G\"{u}ntner,  A. T. and Sievi,  N. A. and Theodore,  S. J. and Gulich,  T. and Kohler,  M. and Pratsinis,  S. E.},
  year = {2017},
  month = sep,
  pages = {10578–10584}
}

@article{hansel1997,
  title = {Energy dependencies of the proton transfer reactions},
  volume = {167–168},
  ISSN = {0168-1176},
  url = {http://dx.doi.org/10.1016/S0168-1176(97)00128-6},
  DOI = {10.1016/s0168-1176(97)00128-6},
  journal = {International Journal of Mass Spectrometry and Ion Processes},
  publisher = {Elsevier BV},
  author = {Hansel,  A. and Singer,  W. and Wisthaler,  A. and Schwarzmann,  M. and Lindinger,  W.},
  year = {1997},
  month = nov,
  pages = {697–703}
}

@article{heaney2022,
  title = {The Impact of a Graded Maximal Exercise Protocol on Exhaled Volatile Organic Compounds: A Pilot Study},
  volume = {27},
  ISSN = {1420-3049},
  url = {http://dx.doi.org/10.3390/molecules27020370},
  DOI = {10.3390/molecules27020370},
  number = {2},
  journal = {Molecules},
  publisher = {MDPI AG},
  author = {Heaney,  Liam M. and Kang,  Shuo and Turner,  Matthew A. and Lindley,  Martin R. and Thomas,  C. L. Paul},
  year = {2022},
  month = jan,
  pages = {370}
}

@article{herbig2008,
  title = {Buffered end-tidal (BET) sampling—a novel method for real-time breath-gas analysis},
  volume = {2},
  ISSN = {1752-7163},
  url = {http://dx.doi.org/10.1088/1752-7155/2/3/037008},
  DOI = {10.1088/1752-7155/2/3/037008},
  number = {3},
  journal = {Journal of Breath Research},
  publisher = {IOP Publishing},
  author = {Herbig,  Jens and Titzmann,  Thorsten and Beauchamp,  Jonathan and Kohl,  Ingrid and Hansel,  Armin},
  year = {2008},
  month = sep,
  pages = {037008}
}

@article{inomata2008,
  title = {Technical Note: Determination of formaldehyde mixing ratios in air with PTR-MS: laboratory experiments and field measurements},
  volume = {8},
  ISSN = {1680-7324},
  url = {http://dx.doi.org/10.5194/acp-8-273-2008},
  DOI = {10.5194/acp-8-273-2008},
  number = {2},
  journal = {Atmospheric Chemistry and Physics},
  publisher = {Copernicus GmbH},
  author = {Inomata,  S. and Tanimoto,  H. and Kameyama,  S. and Tsunogai,  U. and Irie,  H. and Kanaya,  Y. and Wang,  Z.},
  year = {2008},
  month = jan,
  pages = {273–284}
}

@article{kalapos2003,
  title = {On the mammalian acetone metabolism: from chemistry to clinical implications},
  volume = {1621},
  ISSN = {0304-4165},
  url = {http://dx.doi.org/10.1016/S0304-4165(03)00051-5},
  DOI = {10.1016/s0304-4165(03)00051-5},
  number = {2},
  journal = {Biochimica et Biophysica Acta (BBA) - General Subjects},
  publisher = {Elsevier BV},
  author = {Kalapos,  Miklós Péter},
  year = {2003},
  month = may,
  pages = {122–139}
}

@article{kim2020,
  title = {Breath Acetone Measurement-Based Prediction of Exercise-Induced Energy and Substrate Expenditure},
  volume = {20},
  ISSN = {1424-8220},
  url = {http://dx.doi.org/10.3390/s20236878},
  DOI = {10.3390/s20236878},
  number = {23},
  journal = {Sensors},
  publisher = {MDPI AG},
  author = {Kim,  Min Jae and Hong,  Sung Hyun and Cho,  Wonhee and Park,  Dong-Hyuk and Lee,  Eun-Byeol and Song,  Yoonkyung and Choe,  Yong-Sahm and Lee,  Jun Ho and Jang,  Yeonji and Lee,  Wooyoung and Jeon,  Justin Y.},
  year = {2020},
  month = dec,
  pages = {6878}
}

@article{king2009,
  title = {Isoprene and acetone concentration profiles during exercise on an ergometer},
  volume = {3},
  ISSN = {1752-7163},
  url = {http://dx.doi.org/10.1088/1752-7155/3/2/027006},
  DOI = {10.1088/1752-7155/3/2/027006},
  number = {2},
  journal = {Journal of Breath Research},
  publisher = {IOP Publishing},
  author = {King,  J and Kupferthaler,  A and Unterkofler,  K and Koc,  H and Teschl,  S and Teschl,  G and Miekisch,  W and Schubert,  J and Hinterhuber,  H and Amann,  A},
  year = {2009},
  month = jun,
  pages = {027006}
}

@article{koenigstein2020,
  title = {Breath acetone change during aerobic exercise is moderated by cardiorespiratory fitness},
  volume = {15},
  ISSN = {1752-7163},
  url = {http://dx.doi.org/10.1088/1752-7163/abba6c},
  DOI = {10.1088/1752-7163/abba6c},
  number = {1},
  journal = {Journal of Breath Research},
  publisher = {IOP Publishing},
  author = {K\"{o}nigstein,  Karsten and Abegg,  Sebastian and Schorn,  Andrea N and Weber,  Ines C and Derron,  Nina and Krebs,  Andreas and Gerber,  Philipp A and Schmidt-Trucks\"{a}ss,  Arno and G\"{u}ntner,  Andreas T},
  year = {2020},
  month = oct,
  pages = {016006}
}

@article{laffel1999,
  title = {Ketone bodies: a review of physiology,  pathophysiology and application of monitoring to diabetes},
  volume = {15},
  ISSN = {1520-7560},
  url = {http://dx.doi.org/10.1002/(SICI)1520-7560(199911/12)15:6<412::AID-DMRR72>3.0.CO;2-8},
  DOI = {10.1002/(sici)1520-7560(199911/12)15:6<412::aid-dmrr72>3.0.co;2-8},
  number = {6},
  journal = {Diabetes/Metabolism Research and Reviews},
  publisher = {Wiley},
  author = {Laffel,  Lori},
  year = {1999},
  month = nov,
  pages = {412–426}
}

@article{lee2024,
  title = {Breath Analyzer for Real-Time Exercise Fat Burning Prediction: Oral and Alveolar Breath Insights with CNN},
  volume = {10},
  ISSN = {2379-3694},
  url = {http://dx.doi.org/10.1021/acssensors.4c02502},
  DOI = {10.1021/acssensors.4c02502},
  number = {4},
  journal = {ACS Sensors},
  publisher = {American Chemical Society (ACS)},
  author = {Lee,  Byeongju and Lee,  Junyeong and Lee,  Hyung-Kun and Park,  HyungJu and Kwack,  Myung-Joon and Kim,  Do Yeob and Park,  Inkyu and Lim,  Soo and Lee,  Dae-Sik},
  year = {2024},
  month = dec,
  pages = {2510–2519}
}

@inproceedings{liu2009,
  title = {Otsu Method and K-means},
  url = {http://dx.doi.org/10.1109/HIS.2009.74},
  DOI = {10.1109/his.2009.74},
  booktitle = {2009 Ninth International Conference on Hybrid Intelligent Systems},
  publisher = {IEEE},
  author = {Liu,  Dongju and Yu,  Jian},
  year = {2009},
  pages = {344–349}
}

@article{mochalski2024,
  title = {Unravelling the origin of isoprene in the human body—a forty year Odyssey},
  volume = {18},
  ISSN = {1752-7163},
  url = {http://dx.doi.org/10.1088/1752-7163/ad4388},
  DOI = {10.1088/1752-7163/ad4388},
  number = {3},
  journal = {Journal of Breath Research},
  publisher = {IOP Publishing},
  author = {Mochalski,  P and King,  J and Unterkofler,  K and Mayhew,  C A},
  year = {2024},
  month = may,
  pages = {032001}
}

@article{mueller2013,
  title = {A new software tool for the analysis of high resolution PTR-TOF mass spectra},
  volume = {127},
  ISSN = {0169-7439},
  url = {http://dx.doi.org/10.1016/j.chemolab.2013.06.011},
  DOI = {10.1016/j.chemolab.2013.06.011},
  journal = {Chemometrics and Intelligent Laboratory Systems},
  publisher = {Elsevier BV},
  author = {M\"{u}ller,  Markus and Mikoviny,  Tomáš and Jud,  Werner and D’Anna,  Barbara and Wisthaler,  Armin},
  year = {2013},
  month = aug,
  pages = {158–165}
}

@article{nagamine2022,
  title = {Mixed effects of moderate exercise and subsequent various food ingestion on breath acetone},
  volume = {17},
  ISSN = {1752-7163},
  url = {http://dx.doi.org/10.1088/1752-7163/ac9ed4},
  DOI = {10.1088/1752-7163/ac9ed4},
  number = {1},
  journal = {Journal of Breath Research},
  publisher = {IOP Publishing},
  author = {Nagamine,  Koichiro and Mineta,  Daiki and Ishida,  Koji and Katayama,  Keisho and Kondo,  Takaharu},
  year = {2022},
  month = nov,
  pages = {016004}
}

@article{pageglave2018,
  title = {Caloric Expenditure Estimation Differences between an Elliptical Machine and Indirect Calorimetry},
  volume = {2},
  ISSN = {2508-9056},
  url = {http://dx.doi.org/10.26644/em.2018.008},
  DOI = {10.26644/em.2018.008},
  journal = {Exercise Medicine},
  publisher = {Sapientia Publishing Group},
  author = {Glave,  A. Page and Didier,  Jennifer J. and Oden,  Gary L. and Wagner,  Matthew C.},
  year = {2018},
  month = apr,
  pages = {8}
}

@article{prazeller2003,
  title = {Proton transfer reaction ion trap mass spectrometer},
  volume = {17},
  ISSN = {1097-0231},
  url = {http://dx.doi.org/10.1002/rcm.1088},
  DOI = {10.1002/rcm.1088},
  number = {14},
  journal = {Rapid Communications in Mass Spectrometry},
  publisher = {Wiley},
  author = {Prazeller,  Peter and Palmer,  Peter T. and Boscaini,  Elena and Jobson,  Tom and Alexander,  Michael},
  year = {2003},
  month = jun,
  pages = {1593–1599}
}

@article{purdom2018,
  title = {Understanding the factors that effect maximal fat oxidation},
  volume = {15},
  ISSN = {1550-2783},
  url = {http://dx.doi.org/10.1186/s12970-018-0207-1},
  DOI = {10.1186/s12970-018-0207-1},
  number = {1},
  journal = {Journal of the International Society of Sports Nutrition},
  publisher = {Informa UK Limited},
  author = {Purdom,  Troy and Kravitz,  Len and Dokladny,  Karol and Mermier,  Christine},
  year = {2018},
  month = jan 
}

@article{reiner2013,
  title = {Long-term health benefits of physical activity – a systematic review of longitudinal studies},
  volume = {13},
  ISSN = {1471-2458},
  url = {http://dx.doi.org/10.1186/1471-2458-13-813},
  DOI = {10.1186/1471-2458-13-813},
  number = {1},
  journal = {BMC Public Health},
  publisher = {Springer Science and Business Media LLC},
  author = {Reiner,  Miriam and Niermann,  Christina and Jekauc,  Darko and Woll,  Alexander},
  year = {2013},
  month = sep 
}

@article{ruzsanyi2017,
  title = {Breath acetone as a potential marker in clinical practice},
  volume = {11},
  ISSN = {1752-7163},
  url = {http://dx.doi.org/10.1088/1752-7163/aa66d3},
  DOI = {10.1088/1752-7163/aa66d3},
  number = {2},
  journal = {Journal of Breath Research},
  publisher = {IOP Publishing},
  author = {Ruzsányi,  Veronika and Péter Kalapos,  Miklós},
  year = {2017},
  month = jun,
  pages = {024002}
}

@article{schubert2017,
  title = {DBSCAN Revisited,  Revisited: Why and How You Should (Still) Use DBSCAN},
  volume = {42},
  ISSN = {1557-4644},
  url = {http://dx.doi.org/10.1145/3068335},
  DOI = {10.1145/3068335},
  number = {3},
  journal = {ACM Transactions on Database Systems},
  publisher = {Association for Computing Machinery (ACM)},
  author = {Schubert,  Erich and Sander,  J\"{o}rg and Ester,  Martin and Kriegel,  Hans Peter and Xu,  Xiaowei},
  year = {2017},
  month = jul,
  pages = {1–21}
}

@article{shcherbina2017,
  title = {Accuracy in Wrist-Worn,  Sensor-Based Measurements of Heart Rate and Energy Expenditure in a Diverse Cohort},
  volume = {7},
  ISSN = {2075-4426},
  url = {http://dx.doi.org/10.3390/jpm7020003},
  DOI = {10.3390/jpm7020003},
  number = {2},
  journal = {Journal of Personalized Medicine},
  publisher = {MDPI AG},
  author = {Shcherbina,  Anna and Mattsson,  C. and Waggott,  Daryl and Salisbury,  Heidi and Christle,  Jeffrey and Hastie,  Trevor and Wheeler,  Matthew and Ashley,  Euan},
  year = {2017},
  month = may,
  pages = {3}
}

@article{smith2017,
  title = {On the importance of accurate quantification of individual volatile metabolites in exhaled breath},
  volume = {11},
  ISSN = {1752-7163},
  url = {http://dx.doi.org/10.1088/1752-7163/aa7ab5},
  DOI = {10.1088/1752-7163/aa7ab5},
  number = {4},
  journal = {Journal of Breath Research},
  publisher = {IOP Publishing},
  author = {Smith,  David and Španěl,  Patrik},
  year = {2017},
  month = nov,
  pages = {047106}
}

@article{spanel2011,
  title = {Breath acetone concentration; biological variability and the influence of diet},
  volume = {32},
  ISSN = {1361-6579},
  url = {http://dx.doi.org/10.1088/0967-3334/32/8/N01},
  DOI = {10.1088/0967-3334/32/8/n01},
  number = {8},
  journal = {Physiological Measurement},
  publisher = {IOP Publishing},
  author = {Španěl,  Patrik and Dryahina,  Kseniya and Rejšková,  Alžběta and Chippendale,  Thomas W E and Smith,  David},
  year = {2011},
  month = jul,
  pages = {N23–N31}
}

@article{spanel2020,
  title = {Quantification of volatile metabolites in exhaled breath by selected ion flow tube mass spectrometry,  SIFT-MS},
  volume = {16},
  ISSN = {2376-9998},
  url = {http://dx.doi.org/10.1016/j.clinms.2020.02.001},
  DOI = {10.1016/j.clinms.2020.02.001},
  journal = {Clinical Mass Spectrometry},
  publisher = {Stanford University Press},
  author = {Španěl,  Patrik and Smith,  David},
  year = {2020},
  month = apr,
  pages = {18–24}
}

@article{sukul2023,
  title = {Origin of breath isoprene in humans is revealed via multi-omic investigations},
  volume = {6},
  ISSN = {2399-3642},
  url = {http://dx.doi.org/10.1038/s42003-023-05384-y},
  DOI = {10.1038/s42003-023-05384-y},
  number = {1},
  journal = {Communications Biology},
  publisher = {Springer Science and Business Media LLC},
  author = {Sukul,  Pritam and Richter,  Anna and Junghanss,  Christian and Schubert,  Jochen K. and Miekisch,  Wolfram},
  year = {2023},
  month = sep 
}

@article{suntrup2020,
  title = {Characterization of a high-resolution breath acetone meter for ketosis monitoring},
  volume = {8},
  ISSN = {2167-8359},
  url = {http://dx.doi.org/10.7717/peerj.9969},
  DOI = {10.7717/peerj.9969},
  journal = {PeerJ},
  publisher = {PeerJ},
  author = {Suntrup III,  Donald J. and Ratto,  Timothy V. and Ratto,  Matt and McCarter,  James P.},
  year = {2020},
  month = sep,
  pages = {e9969}
}

@article{turner2006,
  title = {A longitudinal study of ammonia,  acetone and propanol in the exhaled breath of 30 subjects using selected ion flow tube mass spectrometry,  SIFT-MS},
  volume = {27},
  ISSN = {1361-6579},
  url = {http://dx.doi.org/10.1088/0967-3334/27/4/001},
  DOI = {10.1088/0967-3334/27/4/001},
  number = {4},
  journal = {Physiological Measurement},
  publisher = {IOP Publishing},
  author = {Turner,  Claire and Španěl,  Patrik and Smith,  David},
  year = {2006},
  month = feb,
  pages = {321–337}
}

@article{wang2013,
  title = {Is breath acetone a biomarker of diabetes? A historical review on breath acetone measurements},
  volume = {7},
  ISSN = {1752-7163},
  url = {http://dx.doi.org/10.1088/1752-7155/7/3/037109},
  DOI = {10.1088/1752-7155/7/3/037109},
  number = {3},
  journal = {Journal of Breath Research},
  publisher = {IOP Publishing},
  author = {Wang,  Zhennan and Wang,  Chuji},
  year = {2013},
  month = aug,
  pages = {037109}
}

@article{warneke2005,
  title = {Development of proton-transfer ion trap-mass spectrometry: on-line detection and identification of volatile organic compounds in air},
  volume = {16},
  ISSN = {1044-0305},
  url = {http://dx.doi.org/10.1016/j.jasms.2005.03.025},
  DOI = {10.1016/j.jasms.2005.03.025},
  number = {8},
  journal = {Journal of the American Society for Mass Spectrometry},
  publisher = {American Chemical Society (ACS)},
  author = {Warneke,  C. and de Gouw,  J. A. and Lovejoy,  E. R. and Murphy,  P. C. and Kuster,  W. C. and Fall,  R.},
  year = {2005},
  month = aug,
  pages = {1316–1324}
}

@article{weber2021,
  title = {Monitoring Lipolysis by Sensing Breath Acetone down to Parts‐per‐Billion},
  volume = {1},
  ISSN = {2688-4046},
  url = {http://dx.doi.org/10.1002/smsc.202100004},
  DOI = {10.1002/smsc.202100004},
  number = {4},
  journal = {Small Science},
  publisher = {Wiley},
  author = {Weber,  Ines C. and Derron,  Nina and K\"{o}nigstein,  Karsten and Gerber,  Philipp A. and G\"{u}ntner,  Andreas T. and Pratsinis,  Sotiris E.},
  year = {2021},
  month = mar 
}

@techreport{who2022obesity,
  author      = {{Nutrition, Physical Activity \& Obesity, Office for Prevention \& Control of NCDs}},
  title       = {WHO European Regional Obesity Report 2022},
  institution = {World Health Organization},
  year        = {2022},
  month       = may,
  day         = 2,
  type        = {Report},
  isbn        = {9789289057738},
  url         = {https://iris.who.int/bitstream/handle/10665/353747/9789289057738-eng.pdf}
}

\end{document}